\documentclass[12pt]{report}
\usepackage{placeins}
\usepackage{fancyhdr}

\fancyheadoffset{0cm}

\fancypagestyle{plain}{%
  \fancyhf{}%
  \fancyhead[R]{\thepage}%
}
\usepackage{times}
\usepackage[letterpaper, left=1in, right=1in, top=1in, bottom=1in]{geometry}
\usepackage{setspace}
\usepackage{verbatimbox}
\usepackage[ruled,vlined]{algorithm2e}
\usepackage[explicit]{titlesec} 
\usepackage[titles]{tocloft} 
\usepackage[backend=bibtex, style=numeric-comp, bibstyle=ieee]{biblatex}
\usepackage[bookmarks=true,hidelinks]{hyperref}
\usepackage[page]{appendix}
\usepackage[normalem]{ulem}
\usepackage{indentfirst}
\usepackage{graphicx}
\usepackage[justification=centering]{caption}
\usepackage{subcaption}
\DeclareCaptionLabelSeparator{period}{.\hspace*{0.5em}}
\usepackage{rotating}
\usepackage{multirow}
\usepackage{tabularx}
\usepackage{makecell}
\usepackage{array}

\usepackage{float}
\usepackage[euler]{textgreek}
\usepackage{gensymb}
\usepackage{textcomp}
\usepackage[fleqn]{amsmath}
\usepackage{hyphenat}
\usepackage[dvipsnames]{xcolor}

\usepackage{longtable}
\usepackage[xindy,acronyms,symbols,nonumberlist]{glossaries}
\usepackage{listings}
\makeglossaries
\setglossarystyle{indexhypergroup}
\setacronymstyle{long-sc-short}

\newacronym{abs}{ABS}{acrylonitrile butadiene styrene}
\newacronym{polybut}{PB}{polybutadiene}
\newacronym{fff}{FFF}{fused filament fabrication}

\newglossaryentry{hion}{
  name = $H+$ ,
  description = hydrogen ion,
  sort = hion
}

\bibliography{references.bib}

\AtEveryBibitem{\clearfield{issn}}
\AtEveryBibitem{\clearlist{issn}}

\AtEveryBibitem{\clearfield{isbn}}
\AtEveryBibitem{\clearlist{isbn}}

\AtEveryBibitem{\clearfield{language}}
\AtEveryBibitem{\clearlist{language}}

\AtEveryBibitem{\clearfield{doi}}
\AtEveryBibitem{\clearlist{doi}}

\AtEveryBibitem{\clearfield{note}}
\AtEveryBibitem{\clearlist{note}}

\AtEveryBibitem{\clearfield{url}}
\AtEveryBibitem{\clearlist{url}}

\AtEveryBibitem{%
  \ifentrytype{online}
    {}
    {\clearfield{urlyear}\clearfield{urlmonth}\clearfield{urlday}}}

\begin{document}
\doublespacing 



\newcommand{\thesisTitle}{WOUND HEALING MODELING USING PARTIAL DIFFERENTIAL EQUATION AND DEEP LEARNING}
\newcommand{\yourName}{Hy Dang}
\newcommand{\yourSchool}{Texas Christian University}
\newcommand{\yourMonth}{December 13}
\newcommand{\yourYear}{2021}
\newcommand{\yourDept}{Department of Mathematics}
\newcommand{\yourAdvisor}{\textbf{Advisor:} Dr. Ken Richardson}

\vspace{10cm}
\begin{titlepage}
\begin{center}
\vspace{100\baselineskip}
\begin{spacing}{2}
{\thesisTitle}
\end{spacing}

\vspace{3\baselineskip}
by\\
{\yourName}\\
\vspace{2\baselineskip}
{\yourSchool}\\
Fort Worth, Texas\\
\yourYear
\end{center}
\end{titlepage}

\currentpdfbookmark{Title Page}{titlePage}  


\pagenumbering{roman}
\addcontentsline{toc}{chapter}{Acknowledgments}
\setcounter{page}{1} 
\clearpage
\begin{centering}
\textbf{ACKNOWLEDGEMENTS}\\
\vspace{\baselineskip}
\end{centering}
The photos of wounds used in the training and testing sets were supplied by the research of Dr. Bert Slade and from Smith \& Nephew, Inc.
\clearpage

\pagenumbering{roman}
\addcontentsline{toc}{chapter}{Abstract}
\setcounter{page}{2}
\clearpage
\begin{centering}
\textbf{\MakeUppercase{ABSTRACT}}\\
\vspace{\baselineskip}
\end{centering}

The process of wound healing has been an active area of research around the world. The problem is the wounds of different patients heal differently. For example, patients with a background of diabetes may have difficulties in healing \cite{article}. By clearly understanding this process, we can determine the type and quantity of medicine to give to patients with varying types of wounds. In this research, we use a variation of the Alternating Direction Implicit method to solve a partial differential equation that models part of the wound healing process. Wound images are used as our dataset that we analyze. To segment the image's wound, we implement deep learning-based models. We show that the combination of a variant of the Alternating Direction Implicit method and Deep Learning provides a reasonably accurate model for the process of wound healing. To the best of our knowledge, this is the first attempt to combine both numerical PDE and deep learning techniques in an automated system to capture the long-term behavior of wound healing.


\renewcommand{\cftchapdotsep}{\cftdotsep}  
\renewcommand{\cftchapfont}{\bfseries}  
\renewcommand{\cftchappagefont}{}  
\renewcommand{\cftchapaftersnum}{.}
\renewcommand{\cftchapafterpnum}{\vskip\baselineskip} 
\renewcommand{\cftsecaftersnum}{.}
\renewcommand{\cftsecafterpnum}{\vskip\baselineskip}  
\renewcommand{\cftsubsecaftersnum}{.}
\renewcommand{\cftsubsecafterpnum}{\vskip\baselineskip} 
\renewcommand{\cftsubsubsecafterpnum}{\vskip\baselineskip} 

\titleformat{\chapter}[display]
{\normalfont\bfseries\filcenter}{\chaptertitlename\ \thechapter}{0pt}{\MakeUppercase{#1}}

\renewcommand\contentsname{Table of Contents}

\begin{singlespace}
\tableofcontents
\end{singlespace}

\currentpdfbookmark{Table of Contents}{TOC}

\addcontentsline{toc}{chapter}{List of Figures}
\begin{singlespace}
\setlength\cftbeforefigskip{\baselineskip}  
\listoffigures
\end{singlespace}



%

%
%


\pagenumbering{arabic}
\setcounter{page}{1} 

\titleformat{\chapter}{\normalfont\bfseries\filcenter}{\MakeUppercase\chaptertitlename\ \thechapter.}{1em}{\MakeUppercase{#1}}  
\titlespacing*{\chapter}
  {0pt}{0pt}{30pt}	
  
\titleformat{\section}{\normalfont\bfseries}{\thesection}{1em}{#1}

\titleformat{\subsection}{\normalfont}{\thesubsection}{1em}{\uline{#1}}

\titleformat{\subsubsection}{\normalfont\itshape}{\thesubsection}{1em}{#1}



\chapter{Introduction}
\label{ch:intro}

\section{Biological Background \& Motivation}

According to \cite{sherratt1990models}, there are three stages of the wound healing process:  inflammation, wound closure, and matrix remodeling in scar tissue.

\begin{figure}[H]
    \centering
    \includegraphics[scale=5]{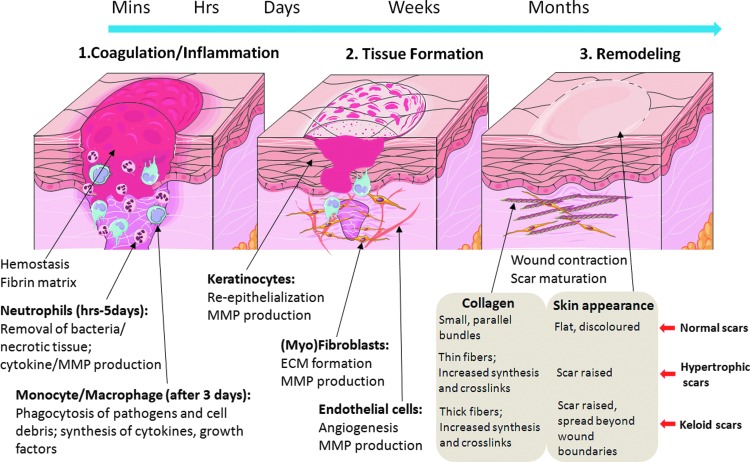}
    \caption{The three classic stages of cutaneous wound healing \cite{xue2015extracellular}}
    \label{fig:example}
\end{figure}
\section{Literature Review}
\subsection{Wound Healing Model Equation}
It is undeniable that researchers want to understand human behavior and themselves. According to \cite{flegg2015mathematical}, various research projects have attempted to present mathematical descriptions of the wound healing process. Due to \cite{sherratt1990models}, a single reaction-diffusion equation is introduced for epidermal cell density. This equation is the initial step for the future development of the mathematical model. The author considers the wound to be two-dimensional since it is reasonable when comparing the dimensions of the wounds to the thickness of the wound. The author proposes the model of the rate of increase of cell density $n$, which includes two parts: cell migration ($cm$) and mitotic generation $(mg)$. The equation is represented as:
\begin{equation*}
    n = cm + mg\\
    cm = D\nabla\cdot \left[(\frac{n}{n_0})^p\nabla n \right]\\
    mg = s*n(1-(\frac{n}{n_0}))\\
\end{equation*}
where n is rate of increase of cell density, $n_0$ is the cell density for skin that is not wounded. 
\subsection{Bio-medical Segmentation}
To apply wound analysis, we need to get information about the wound, including wound areas, wound boundaries, etc. However, due to the high volume of patients, time-consuming quantitative wound assessments are not always feasible \cite{7318881}.
With the development of big data and technology, traditional data processing can not analyze information efficiently \cite{wang2020recent}. 
\section{Motivation}
We need a mathematical model to better understand the wound healing process. We believe that the healing process follows specific stages. Moreover, by modeling the process mathematically, we could determine the appropriate treatment for patients. We are trying to build and construct an end-to-end process of wound healing. Moreover, we want to reduce manual, error-prone steps in our process, and we will utilize the advantages of machine learning to efficiently analyze healing. 
\section{Implementation}
Our purpose is to build a system that uses Deep Learning and a variant of the Alternating Direction Implicit method. Thus, our structure combines three primary steps including \textbf{Segmentation}, \textbf{Homothety} and \textbf{Predicting Wound Behavior}.
In more detail, we first need a stage that could capture all the helpful data from the datasets (which are in image format). In more general words, we need to identify and segment the wounds from the images. Then, we want to scale, rotate, and translate all the images of patients in different times to the same scale and orientation. After preparing the dataset, we want to apply these initial conditions to a partial differential equation (PDE) model to predict the wound behavior.


\chapter{METHODS}
\label{ch:methods}
\section{Deep Learning}
\label{sect:deeplearning}
\subsection{Introduction}
\label{sect:dl:intro}

Wound segmentation plays an important role in our process to label parts of the skin and the background in each image. However, limitations and time constraints of manual labelling make data processing not feasible. According to \cite{li2018composite}, the manual methods are time-consuming. Moreover, the outcomes are not accurate.

We want to exploit the advantages of Deep Learning, which has been popular in recent years. It uses the powerful feature extraction ability of the trained deep neural networks to process images \cite{li2018composite}. Due to its achievement in computer vision problems including image classification, recognition, and segmentation, we believe that deep learning can handle the segmentation problem in our project. 
\newpage
\subsection{U-Net Model}
\begin{figure}[!htbp]
    \centering
    \includegraphics[scale = 0.3]{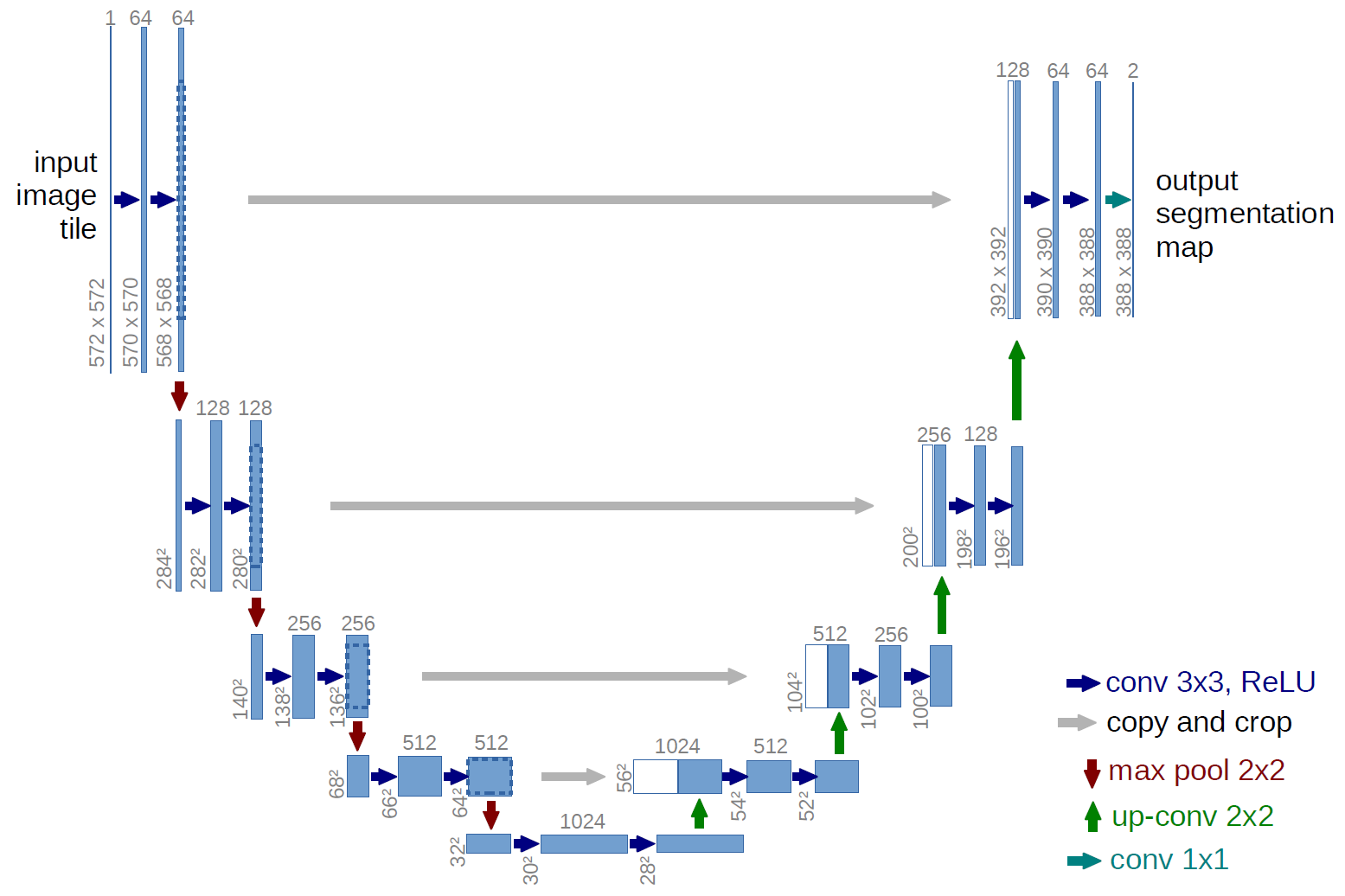}
    \caption{U-net architecture \cite{RFB15a}}
    \label{fig:u-net}
\end{figure}

The U-Net is an architecture with precise segmentation of images \cite{RFB15a} (see Figure~\ref{fig:u-net}). The important feature of this model is when it only requires a few training samples and can generate precise segmentation of images. Thus, the model can resolve the problem of limited training data. Moreover, U-Net has proved good performance on biomedical segmentation tasks, which showed a potential approach for our research problem. 
\FloatBarrier
\subsection{Residual U-Net Model}
\begin{figure}[!htbp]
\centering
\begin{minipage}{.5\textwidth}
  \centering
  \includegraphics[width=.9\linewidth]{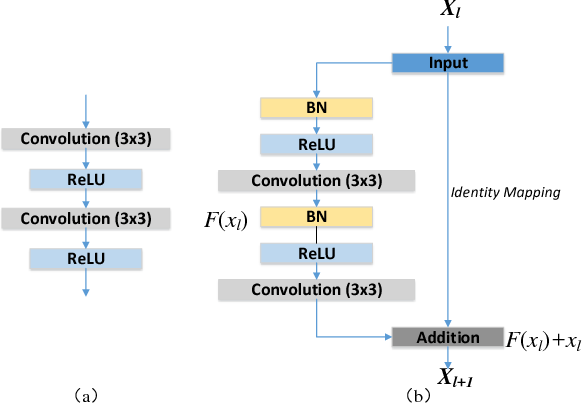}
    \captionof{figure}{Building blocks of neural networks \cite{zhang2018road}}
    \label{fig:block_unet}
\end{minipage}%
\begin{minipage}{.5\textwidth}
  \centering
  \includegraphics[width=.7\linewidth]{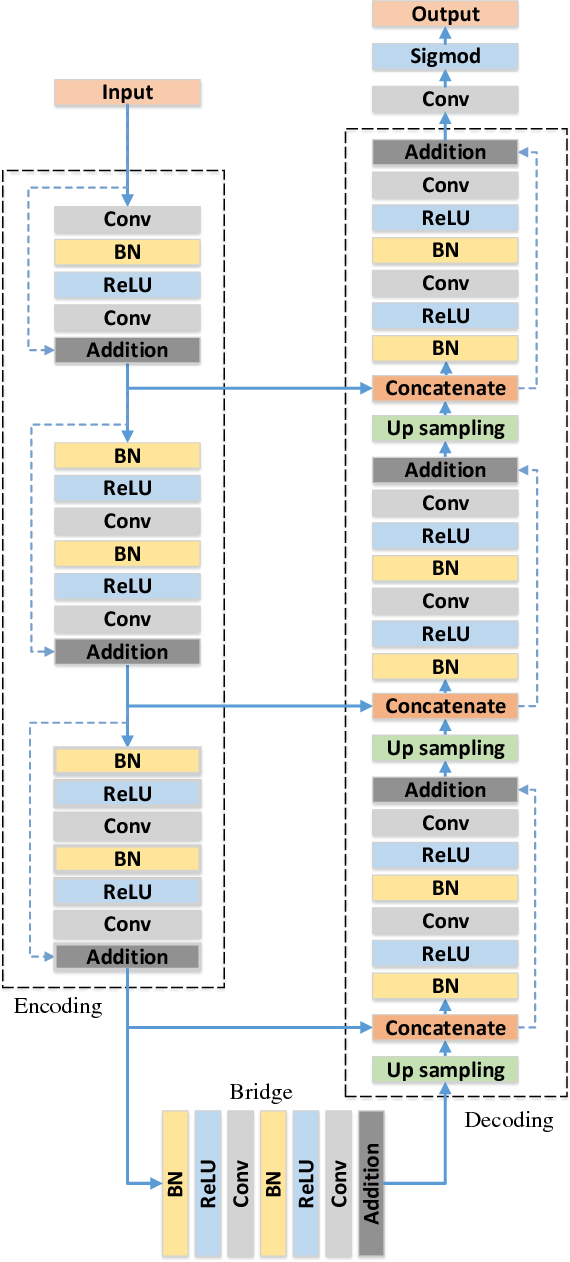}
  \captionof{figure}{The architecture of the proposed deep ResUnet. \cite{zhang2018road}}
  \label{fig:resunet}
\end{minipage}
\end{figure}

Zhang et al. proposed the Residual U-Net (see Figure~\ref{fig:resunet}), a neural network combining residual units associated with the mentioned U-Net model. According to \cite{zhang2018road}, the model brings two main benefits. The first advantage is the ease of training of deep networks caused by the residual units. Moreover, the skip connections in the architecture contribute to better performance with fewer parameters. The architecture proved its efficiency in road-extraction problems and outperformed the U-Net model in this category. Therefore, we adopted this architecture to validate our segmentation task for the wound-segmentation problem.
\subsection{Mask-RCNN}
\label{mask-rcnn}
Mask-RCNN has been the latest state-of-the-art method of instance segmentation. Mask-RCNN includes two stages. First, it identifies bounding boxes of object candidates. This phase is referred to as Region Proposal Network (RPN) in Faster-RCNN architecture \cite{ren2016faster}. The second stage of Mask-RCNN also generates a binary mask for each RoI apart from predicting the class of the object and refining the box \cite{he2017mask}. 
Detectron2 is Facebook AI Research's next-generation library that uses Mask-RCNN and provides state-of-the-art detection and segmentation algorithms.
\begin{figure}[!htbp]
    \centering
    \includegraphics[scale = 0.6]{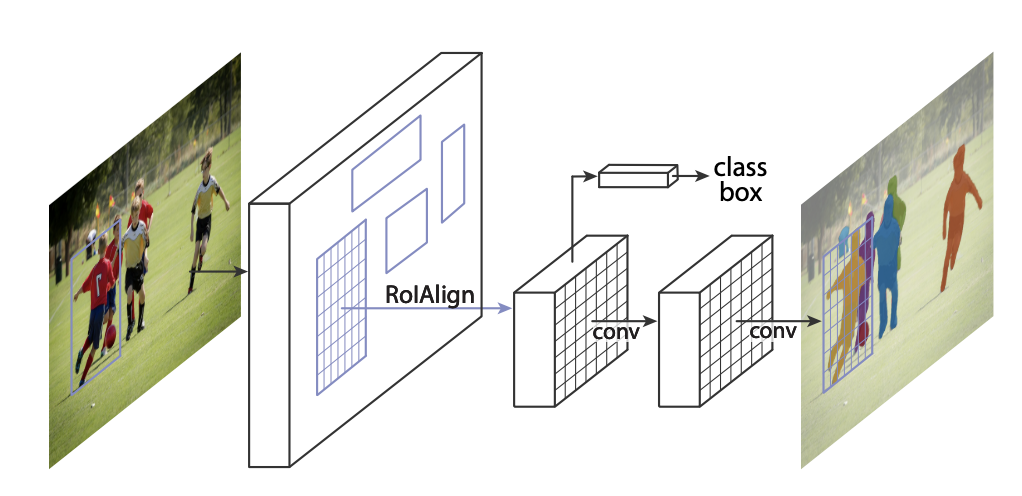}
    \caption{The Mask R-CNN framework for instance segmentation \cite{he2017mask}}
    \label{fig:mask_rcnn}
\end{figure}
\FloatBarrier
\section{Homothety: Rescaling And Mapping Images}
\subsection{Problem}
Since the dataset images were taken in different scales and angles, it's difficult for the PDE Approximation Methods to work with the images at later times. 
\subsection{Solution}
Thus, we need to find mapping methods which align the images. We go through different stages to handle this problem. We proposed an alignment method which we call \textbf{Mappler}. 
\subsection{Mappler}
\label{mappler}
\begin{algorithm}[H]
\SetAlgoLined
\KwResult{List of tuples (Angle Rotation, Zoom Scaling)}
    initialization:
    pick the first image as the main image\\
    convert all images to gray scale channel\;
    \For{image}{
        getting center of mass of the wound\; 
        getting smaller window of image (from center of mass)\;
        mapping centers of mass between image and main image\;
        consider matched pixels \;
        \For{(rotation, zoom scale)}{
        dictionary $\gets$ compute number of matched pixels}
    }
    return value with maximum number of matches in dictionary

\caption{Mappler}
\end{algorithm}

\section{Diffusion Equation Approximation}
\subsection{Overview - Partial Differential Equations}
\indent A \textbf{partial differential equation} (PDE) is a differential equation that includes multivariable functions and their partial derivatives. These functions are adopted to formulate problems with functions of several variables. There are analytic and numerical approaches to solve PDEs.
While ordinary differential equations (ODE) often be used to model one-dimensional dynamical systems, PDEs usually model multidimensional ones. These have been an attractive field for scientific discoveries where PDEs are implemented to represent many systems and structures from physics, chemistry, computer science, etc.

An initial value problem (IVP) for two-dimensional heat equation on a square of side length $L$ has the form:
\begin{eqnarray}
\frac{\partial u}{\partial t} &=& k\left(\frac{\partial ^ 2 u}{\partial x^2}+\frac{\partial ^ 2 u}{\partial y^2}\right); \hspace{0.2cm} t > 0 ; \hspace{0.2cm} (x,y) \in [0,L] \times [0,L]; \nonumber\\
u(x,y,0) &=& I(x,y) 
\label{IVP}
\end{eqnarray}

\subsection{Forward Euler Method}
\label{euler}
First, we try to numerically solve the IVP by using the Euler Method, which is straightforward. Euler's method is based on the approximations of the graph of a solution by approximating tangent line sequentially in steps \cite{howell2019ordinary}.

The grid on $\omega = [0,L_x] \times [0,L_y]$ with grid point $(x_i,y_j)$ is introduced where $x_i = i\Delta x$ and $y_j = j\Delta y$, for $i = 0,...,n_x, j= 0,...,n_y.$

From (\ref{IVP}), we have: 
\begin{eqnarray*}
\frac{u_{i,j}^{n+1} - u_{i,j}^{n}}{\Delta t} &=& \frac{u_{i-1,j}^n-2u_{i,j}^n+u_{i+1,j}^n}{\Delta x^2} + \frac{u_{i,j-1}^n-2u_{i,j}^n+u_{i,j+1}^n}{\Delta y^2} \\
\Longleftrightarrow{} u_{i,j}^{n+1} &=& \frac{\Delta t}{\Delta x^2}(u_{i-1,j}^n-2u_{i,j}^n+u_{i+1,j}^n) + \frac{\Delta t}{\Delta y^2}(u_{i,j-1}^n-2u_{i,j}^n+u_{i,j+1}^n) + u_{i,j}^{n}
\\&=:&Euler(u^n, i, j)
\end{eqnarray*}

The pseudocode for the Forward Euler Method is: 
\definecolor{codegreen}{rgb}{0,0.6,0}
\definecolor{codegray}{rgb}{0.5,0.5,0.5}
\definecolor{codepurple}{rgb}{0.58,0,0.82}
\definecolor{backcolour}{rgb}{0.95,0.95,0.92}
 
\lstdefinestyle{mystyle}{
    backgroundcolor=\color{backcolour},   
    commentstyle=\color{codegreen},
    keywordstyle=\color{magenta},
    numberstyle=\tiny\color{codegray},
    stringstyle=\color{codepurple},
    basicstyle=\ttfamily\footnotesize,
    breakatwhitespace=false,         
    breaklines=true,                 
    captionpos=b,                    
    keepspaces=true,                 
    numbers=left,                    
    numbersep=5pt,                  
    showspaces=false,                
    showstringspaces=false,
    showtabs=false,                  
    tabsize=2
}
\begin{algorithm}[H]
\SetAlgoLined
\KwResult{Wound Solution}
    \textbf{Parameters:}\\
    - \verb|L|: the end interval of x \\
    - \verb|Nt, Nx, Ny:| the number of mesh points in time and in x, y \\
    - \verb|T|: the stop time for simulation\\
    - \verb|I|: the initial condition of the function u(0,x,y)\\
    - \verb|u = u_n = [Nx+1, Ny+1]|\\
    - \verb|u[0,i,j] = I[i,j] for i = 0,1,..., k_x; j = 0,1,..., k_y| \\
    \begin{verbatim}
for n in (0,Nt) do
    for i in (0,Nx) do
        for j in (0,Ny) do
            u[n+1,i,j] = Euler(u^n,i,j)
    \end{verbatim}
\caption{Euler Method}
\end{algorithm}
\vspace{0.2in}
To test the Euler method, we implement the function where we know the exact solution.
\begin{eqnarray}
       f(x,y,t) &=& \sin(x) \sin(y) e^{-2t}  
\end{eqnarray}
Then, for input \verb|L = math.pi, dt = 0.1, dx,dy = 0.05,0.05, T = 0.2|
\begin{figure}
\centering
\begin{minipage}{.5\textwidth}
  \centering
  \includegraphics[width=1\linewidth]{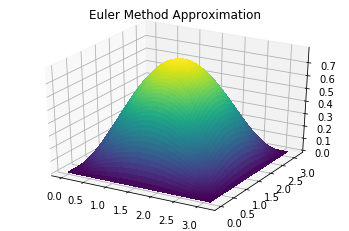}
  \label{fig:sub1}
\end{minipage}%
\begin{minipage}{.5\textwidth}
  \centering
  \includegraphics[width=1\linewidth]{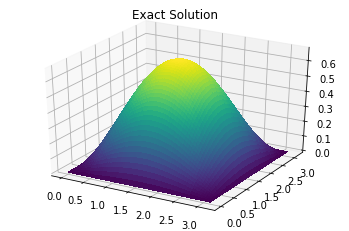}
  \label{fig:sub2}
\end{minipage}
\label{fig:test}
\caption{Euler approximation with $L = \pi, dt = 0.1, dx = dy = 0.05, T = 0.2$}
\end{figure}
\\
It is easy to see that there is a good result for the \textbf{Euler Method Approximation}. However, we want to determine the accuracy numerically. For this reason, the \textbf{Mean Square Error} is used:
\begin{eqnarray}
MSE = \frac{1}{n}\sum_{i=1}^{n}{\left(\frac{d_i -f_i}{\sigma_i}\right)^2}
\end{eqnarray}
And the $MSE$ for \textbf{Euler Method Approximation} is $16.697$ for all of the above inputs.
\subsection{Alternating Method (ADI Method)}
The Euler method presented in the previous section is flexible and efficient, but not the optimal solution for the two-dimensional heat equation. Instead, the ADI method is introduced with the idea of replacing a two-dimensional problem with a series of one-dimensional problems to generate a computationally efficient algorithm. For this method, we also have the same grid $\omega$ as in the above Euler Method. However, we then divide each time step into two steps of size $\Delta t/2$ and in each substep, the solution is computed in one dimension.
\begin{eqnarray}
\frac{u_{i,j}^{n+\frac{1}{2}} - u_{i,j}^{n}}{\Delta t/2} &=& \frac{u_{i-1,j}^{n+\frac{1}{2}}-2u_{i,j}^{n+\frac{1}{2}}+u_{i+1,j}^{n+\frac{1}{2}}}{\Delta x^2} + \frac{u_{i,j-1}^n-u_{i,j}^n+u_{i,j+1}^n}{\Delta y^2} \nonumber
\\
\Longleftrightarrow{} -\frac{\Delta t}{\Delta x^2}u_{i-1,j}^{n+\frac{1}{2}}&+&2(1+\frac{\Delta t}{2\Delta x^2})u_{i,j}^{n+\frac{1}{2}} - \frac{\Delta t}{2\Delta x^2} u_{i+1,j}^{n+\frac{1}{2}} = 2u_{i,j}^n + \frac{\Delta t}{2\Delta y^2}(u_{i,j-1}^n - 2u_{i,j}^n) \nonumber\\
&:=& ADI_1(u[n,i,j]) 
\label{first_adi}
\end{eqnarray}
Similarly, we have for another sub time step: 
\begin{eqnarray}
\frac{u_{i,j}^{n+1} - u_{i,j}^{n+\frac{1}{2}}}{\Delta t/2} &=& \frac{u_{i-1,j}^{n+\frac{1}{2}}-2u_{i,j}^{n+\frac{1}{2}}+u_{i+1,j}^{n+\frac{1}{2}}}{\Delta x^2} + \frac{u_{i,j-1}^{n+1}-u_{i,j}^{n+1}+u_{i,j+1}^{n+1}}{\Delta y^2} \nonumber \\
\Longleftrightarrow{} -\frac{\Delta t}{\Delta y^2}u_{i,j-1}^{n+1}&+&2(1+\frac{\Delta t}{2\Delta y^2})u_{i,j}^{n+1} - \frac{\Delta t}{2\Delta y^2} u_{i,j+1}^{n+1} = 2u_{i,j}^{n+\frac{1}{2}} + \frac{\Delta t}{2\Delta y^2}(u_{i,j-1}^{n+\frac{1}{2}} - 2u_{i,j}^{n+\frac{1}{2}})
\nonumber\\
&:=& ADI_2(u[n+1/2,i,j]) 
\label{second_adi}
\end{eqnarray}
By writing (\ref{first_adi}), (\ref{second_adi}) in matrix form, we have:
\[
\begin{bmatrix}
1 & & & & &  \\
-\frac{\Delta t}{2\Delta x^2} & 2 + 2\frac{\Delta t}{2\Delta x^2} & -\frac{\Delta t}{2\Delta x^2}&&&\\
&-\frac{\Delta t}{2\Delta x^2} & 2 + 2\frac{\Delta t}{2\Delta x^2} & -\frac{\Delta t}{2\Delta x^2}&&\\
&&\ddots&\ddots&\ddots&\\
&&&-\frac{\Delta t}{2\Delta x^2} & 2 + 2\frac{\Delta t}{2\Delta x^2} & -\frac{\Delta t}{2\Delta x^2}\\
& & & & & 1
\end{bmatrix} 
\times 
\left[ \begin{array}{c} u_{0,j}^{n+\frac{1}{2}} \\ u_{1,j}^{n+\frac{1}{2}}\\u_{2,j}^{n+\frac{1}{2}}\\\vdots\\u_{k_{x-1},j}^{n+\frac{1}{2}}\\u_{k_{x},j}^{n+\frac{1}{2}} \end{array} \right]  
= \left[\begin{array}{c} b_{0,j}\\b_{1,j}\\b_{2,j}\\\vdots\\b_{{k_x-1},j}\\b_{{k_x},j} \end{array}\right]
\]
and
\[
\begin{bmatrix}
1 & & & & &  \\
-\frac{\Delta t}{2\Delta y^2} & 2 + 2\frac{\Delta t}{2\Delta y^2} & -\frac{\Delta t}{2\Delta y^2}&&&\\
&-\frac{\Delta t}{2\Delta y^2} & 2 + 2\frac{\Delta t}{2\Delta y^2} & -\frac{\Delta t}{2\Delta y^2}&&\\
&&\ddots&\ddots&\ddots&\\
&&&-\frac{\Delta t}{2\Delta y^2} & 2 + 2\frac{\Delta t}{2\Delta y^2} & -\frac{\Delta t}{2\Delta y^2}\\
& & & & & 1
\end{bmatrix} 
\times 
\left[ \begin{array}{c} u_{i,0}^{n+1} \\ u_{i,1}^{n+1}\\u_{i,2}^{n}\\\vdots\\u_{i,k_{y-1}}^{n}\\u_{i,k_{y}}^{n} \end{array} \right]  = \left[\begin{array}{c} c_{i,0}\\c_{i,1}\\c_{i,2}\\\vdots\\c_{i,k_{y-1}}\\c_{i,k_{y}} \end{array}\right].
\]

The steps of the procedure are: First, we solve the 
$k_y - 2$ systems of equations on the first form to obtain the values for the intermediate solution 
$u_{i,j}^{n+\frac{1}{2}}$. 
Then we solve $k_x - 2$ 
systems of equations on the form of the next equation to compute the values $u_{i,j}^{n+1}$ 
from the computed values 
$u_{i,j}^{n+\frac{1}{2}}$.

The algorithm for this procedure is: \\
\begin{algorithm}[H]
\SetAlgoLined
\KwResult{Wound Behavior}
    initialization:\\
    \textbf{Parameters:}\\
    - \verb|L|: the end interval of x \\
    - \verb|Nt, Nx, Ny:| the number of mesh points in time and in x, y \\
    - \verb|T|: the stop time for simulation\\
    - \verb|I|: the initial condition of the function u(0,x)\\
    - \verb|size(u) = [Nt, Nx+1, Ny+1]|\\
    - \verb|u[0,i,j] = I[i,j] for i = 0,1,..., k_x; j = 0,1,..., k_y|\\
    \begin{verbatim}
for n in (1,Nt)
    for i in (1,Nx)
        u[n+1/2,i,j] = ADI_1(u[n,i,j])
        for i in (1,Ny)
            u[n+1,i,j] = ADI_2(u[n+1/2,i,j])
    \end{verbatim}

\caption{Alternating Direction Implicit Method}
\end{algorithm}
We now use the same initial condition from the previous test of the Euler Method to check the accuracy of the ADI method.

\begin{figure}[H]
\centering
\begin{minipage}{.5\textwidth}
  \centering
  \includegraphics[width=1\linewidth]{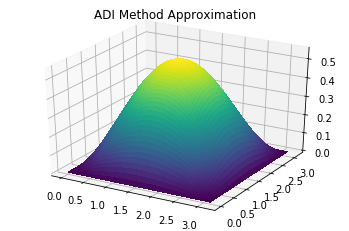}
  \label{fig:adi_error}
\end{minipage}%
\begin{minipage}{.5\textwidth}
  \centering
  \includegraphics[width=1\linewidth]{exact.png}
  \label{fig:exact_fig}
\end{minipage}
\label{fig:test_23}
\caption{ADI approximation with $L = \pi, dt = 0.1, dx = dy = 0.05, T = 0.2$}
\end{figure}

The $MSE$ for the \textbf{ADI Method} is $14.696$ which is less than that of the Euler Method ($16.697$).
For another example, if we choose $dt = 0.02, dx = dy = 0.1$, then the Euler Method gives $MSE = 534119768257.6423$, while the ADI Method still has a good approximation with $MSE = 2.1$.

\begin{figure}
\centering
\begin{minipage}{.45\textwidth}
  \centering
  \includegraphics[width=0.8\linewidth]{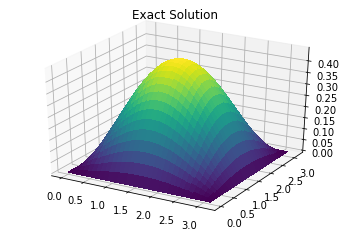}
  \label{fig:sub3}
\end{minipage}%
\begin{minipage}{.45\textwidth}
  \centering
  \includegraphics[width=0.8\linewidth]{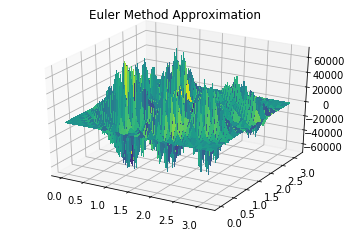}
  \label{fig:sub23}
\end{minipage}
\begin{minipage}{.45\textwidth}
  \centering
  \includegraphics[width=0.8\linewidth]{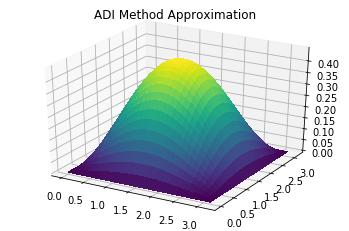}
  \label{fig:sub24}
\end{minipage}
\label{fig:test_25}
\caption{Euler vs ADI approximation with $L = \pi, dt = 0.02, dx = dy = 0.1, T = 0.2$ }
\end{figure}

\subsection{Stability}
We now discuss the stability of both methods, since it was shown by example that the Euler Method sometimes fails to converge to the true solution. The stability of Euler Method depends on the ratio $\frac{\Delta t}{\Delta x^2}$ and $\frac{\Delta t}{\Delta y^2}$. From the equation in Section~\ref{euler}:
\begin{eqnarray*}
u_{i,j}^{n+1} &=& \frac{\Delta t}{\Delta x^2}(u_{i-1,j}^n-2u_{i,j}^n+u_{i+1,j}^n) + \frac{\Delta t}{\Delta y^2}(u_{i,j-1}^n-2u_{i,j}^n+u_{i,j+1}^n) + u_{i,j}^{n}\\
\Longleftrightarrow u_{i,j}^{n+1} &=& \frac{\Delta t}{\Delta x^2}(u_{i-1,j}^n +u_{i+1,j}^n) + \frac{\Delta t}{\Delta y^2}(u_{i,j-1}^n+u_{i,j+1}^n) + u_{i,j}^{n}(1 - 2\frac{\Delta t}{\Delta x^2} - 2\frac{\Delta t}{\Delta y^2}) 
\end{eqnarray*}
So to make the approximation of solution stable, we must make the term $(1 - 2\frac{\Delta t}{\Delta x^2} - 2\frac{\Delta t}{\Delta y^2})$ positive or $\frac{\Delta t}{\Delta x^2} + \frac{\Delta t}{\Delta y^2} \leq \frac{1}{2}$. So for our problem, if we set $\frac{\Delta t}{\Delta x^2} = \frac{\Delta t}{\Delta y^2}$, so that $\frac{\Delta t}{\Delta x^2} \leq \frac{1}{4}$\\
To test our method and stability condition, we consider different values of $\Delta y$ and $\Delta x$, while keeping $\Delta t$ constant and vice versa. The table below returns MSE for Euler and ADI Approximation methods for different situations. 
\begin{table}[!htbp]
    \centering
 \begin{tabular}{||c| c| c| c||} 
 \hline
$\Delta x$ & $\frac{\Delta t}{\Delta x^2}$ & Euler Method's MSE & ADI Method's MSE \\ [0.5ex] 
 \hline
 \hline
 0.5 & 0.04 & 3.46e-05 & 1.23e-05 \\ 
  \hline
 0.30  & 0.11 & 1.47e-05 & 1.80e-06 \\ 
  \hline
 0.20  & 0.25 & 9.74e-06 & 2.94e-07 \\ 
  \hline
 0.13 & 0.592 & 2.16e+08 & 6.0e-08 \\
  \hline
 0.10 & 1.00 & 8.34e+28 & 2.18e-08 \\
 \hline
 0.05 & 4.00 & 8.16e+80 & 1.23e-09 \\ 
\hline
\end{tabular}
\caption{ADI and Euler approximation MSE with $L = \pi, \Delta t = 0.01$ and changing $\Delta x, \Delta y$}
    \label{table:my_label}
\end{table}
\FloatBarrier
\begin{table}[!htbp]
    \centering
 \begin{tabular}{||c | c | c| c||} 
 \hline
 $\Delta t$ & $\frac{\Delta t}{\Delta x^2}$ & Euler Method's MSE & ADI Method's MSE \\ [0.5ex] 
 \hline
  \hline
 0.001 & 0.1 & 1.686e-07 & 2.22e-08 \\
   \hline
 0.003  & 0.3 & 0.005 & 5.18e-07 \\ 
    \hline
 0.006  & 0.6 & 4.95e+38 & 5.19e-07 \\
   \hline
 0.01  & 1 & 8.346e+28 & 2.18e-08 \\ 
  \hline
   0.015 & 1.5 & 8.89e+17 & 4.078e-06 \\ 
 \hline
 0.02 & 2 & 5.2e+08 & 2.05e-08 \\ 
\hline
\end{tabular}
\caption{ADI and Euler approximation MSE with $L = \pi, \Delta x = \Delta y = 0.1$ and changing $\Delta t$}
    \label{tab:my_label}
\end{table}
\FloatBarrier
From the experiment, we can easily see that the Euler Method produces large MSE when the ratio of $\frac{\Delta t}{\Delta x^2} \geq 0.25$ \\
The \textbf{ADI Method} is a \textit{predictor-corrector} scheme where part of the difference operator is implicit in the prediction step or initial step and the other part is implicit in the correction step or final step \cite{hasnain2018finite}. Moreover, because the initial step includes solving for a time step $\frac{\Delta t}{2}$ using the \textit{backward Euler Method} for the $x$ derivative terms and the \textit{forward Euler Method} for the $y$ derivative terms. 
\begin{eqnarray}
\label{adi1}
\frac{u_{i,j}^{n+\frac{1}{2}} - u_{i,j}^{n}}{\Delta t/2} &=& \frac{u_{i-1,j}^{n+\frac{1}{2}}-2u_{i,j}^{n+\frac{1}{2}}+u_{i+1,j}^{n+\frac{1}{2}}}{\Delta x^2} + \frac{u_{i,j-1}^n-2u_{i,j}^n+u_{i,j+1}^n}{\Delta y^2} 
\end{eqnarray}
The final step completes the solution process for a time step by using the \textit{forward Euler Method} for the $x$ derivative term and \textit{backward Euler Method} for the $y$ derivative term.
\begin{eqnarray}
\label{adi2}
\frac{u_{i,j}^{n+1} - u_{i,j}^{n+\frac{1}{2}}}{\Delta t/2} &=& \frac{u_{i-1,j}^{n+\frac{1}{2}}-2u_{i,j}^{n+\frac{1}{2}}+u_{i+1,j}^{n+\frac{1}{2}}}{\Delta x^2} + \frac{u_{i,j-1}^{n+1}-2u_{i,j}^{n+1}+u_{i,j+1}^{n+1}}{\Delta y^2}
\end{eqnarray}
By subtracting \eqref{adi1} and \eqref{adi2} we easily have that
\begin{eqnarray}
\frac{u_{i,j}^{n+\frac{1}{2}} - u_{i,j}^{n} - u_{i,j}^{n+1} + u_{i,j}^{n+\frac{1}{2}}}{\Delta t/2} &=& \frac{u_{i,j-1}^n-2u_{i,j}^n+u_{i,j+1}^n}{\Delta y^2} - \frac{u_{i,j-1}^{n+1}-2u_{i,j}^{n+1}+u_{i,j+1}^{n+1}}{\Delta y^2} \label{adi3} \\
\Longleftrightarrow 2u_{i,j}^{n+\frac{1}{2}} &=& u^n_{i,j} + u^{n+1}_{i,j} \nonumber\\&+& \frac{\Delta t}{2\Delta y^2}(u_{i,j-1}^n-2u_{i,j}^n+u_{i,j+1}^n - u_{i,j-1}^{n+1}+2u_{i,j}^{n+1}-u_{i,j+1}^{n+1}). \nonumber
\end{eqnarray}
Also, by adding \eqref{adi1} and \eqref{adi2} we also get 
\begin{eqnarray}
u_{i,j}^{n+1} - u_{i,j}^n = \frac{\Delta t}{2\Delta x^2} (u_{i-1,j}^{n+\frac{1}{2}}-2u_{i,j}^{n+\frac{1}{2}}+u_{i+1,j}^{n+\frac{1}{2}} + u_{i-1,j}^{n+\frac{1}{2}}-2u_{i,j}^{n+\frac{1}{2}}+u_{i+1,j}^{n+\frac{1}{2}}) + \label{adi4}\\  \frac{\Delta t}{2\Delta y^2}(u_{i,j-1}^n-2u_{i,j}^n+u_{i,j+1}^n + u_{i,j-1}^{n+1}-2u_{i,j}^{n+1}+u_{i,j+1}^{n+1}). \nonumber
\end{eqnarray}
Then, subtracting \eqref{adi4} from \eqref{adi2}, we have that, letting $\delta_x^2u^n_{i,j} = u^n_{i-1,j} - 2u^n_{i,j}+ u^n_{i+1,j}$ and similarly $\delta_y^2u^n_{i,j} = u^n_{i,j-1} - 2u^n_{i,j}+ u^n_{i,j+1}$, 
\begin{eqnarray}
\label{adi:final}
u_{i,j}^{n+1} -  u_{i,j}^n &=& \frac{1}{2}(\frac{\Delta t}{\Delta x^2}\delta_x^2 + \frac{\Delta t}{\Delta y^2}\delta_y^2)(u_{i,j}^{n+1} + u^n_{i,j}) - \frac{\Delta t \Delta t}{4\Delta x^2 \Delta y^2}\delta^2_x \delta^2_y(u_{i,j}^{n+1} - u_{i,j}^n).
\end{eqnarray}
By considering \eqref{adi:final} and the Crank-Nicolson method, from \cite{crank1947practical}, the local discretization error of the ADI method is 
\begin{eqnarray}
O(\Delta x^2)+ O(\Delta y^2) + O(\Delta t^2).
\end{eqnarray}
The full analysis is in Appendix \ref{app:app-A}.

\chapter{Model Implementation}
\label{ch:exp}
\section{Combination of Deep Learning and Modified Alternating Direction Implicit Method}
\begin{figure}[H]
    \centering
    \includegraphics[scale = 0.5]{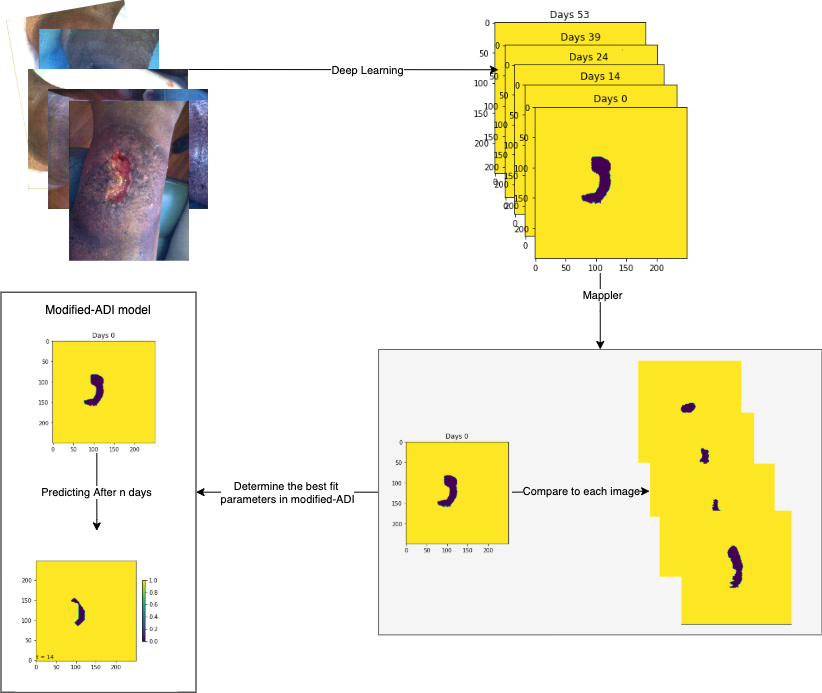}
    \caption{The Process of The Project.}
    \label{fig:combine}
\end{figure}

The process of the project involves different stages and methods, which were mentioned in Chapter \ref{ch:methods}. Firstly, the inputs of the project are wound images from patients at different times. Thus, we need to use the Deep Learning method to segment the wounds in images. We will create two labels for processed images corresponding to the normal skin and the wounded skin. We will examine different Deep Learning algorithms' efficiencies later in Chapter \ref{ch:results}. The Mask-RCNN (Section \ref{mask-rcnn}) achieves the best results among examined models. Thus, we used it as the baseline model for the Deep Learning method. After receiving the predicted images of the wounds from an individual patient, we have to deal with a new problem, which involves the scale and alignment of the images. The images were taken at different times and angles, and the scales of the wounds are different. For this reason, we use the Mappler Algorithm (Section \ref{mappler}) to align images on the same scale. After that, we implement a modified-ADI algorithm to model the progress of the wound. For each patient, we fit the model's parameters to the first four images from the first four recorded days for the wound. We use the subsequent images to test the accuracy of the model. For example, with the patient indicated in Figure~\ref{fig:combine}, we select images from day 0, day 14, day 24, day 39 to fix the model's parameters and the data on day 53 for examining the accuracy of the model. An example of the workflow is represented in Figure~\ref{fig:combine}.
\section{Parameters}
\label{param}
In the modified-ADI method, we use the following parameters to optimize the fit of the model to each individual patient.
\begin{itemize}
    \item $\theta$: This parameter serves as a threshold parameter. It defines the threshold for the pixel in the image to completely healed. For example, suppose a pixel in image after the modified-ADI method returns 0.6. If $\theta = 0.5$. Then that pixel is considered healed skin and labeled appropriately. This parameter $\theta$ is in the range $\left(0.0, 1.0\right)$. After each step of the ADI method, the initial conditions are reset with the value of the function at each pixel taking the value of 0 or 1 according to whether the computed value is $< \theta$ or $\ge \theta$, respectively. This renormalization is not present in the standard ADI method; for this reason, we have named this part of the algorithm the \textit{modified-ADI method}.
    \item $k$: This parameter serves as a speed parameter. It defines the speed of the wound healing process. It serves as the constant in our equation \eqref{IVP}; larger values of $k$ result in faster time evolution.
    \item $\Delta x, \Delta y$: These are the resolution parameters. In our case, since we want to consider each pixel, then $\Delta x = \Delta y = 1$.
    \item $\Delta t$: This is the time-step parameter in \eqref{first_adi} and \eqref{second_adi}.
\end{itemize}
\section{Parameter Searching}
\label{parametersearching}
We use the data from the first four images from each individual to find the best parameters for our model. We will use the loss function that calculates the differences in the area of the wounds between predicted images and the actual images to examine the accuracy of the parameters.
\begin{equation}
    \operatorname{loss} = \frac{\sum_i^n{\text{predicted wound area}_i - \text{actual wound area}_i}}{n},
\end{equation} 
where $n$ is the number of images considered.
Through our experiment, we realize that we can divide our process into 3 main steps. First, we search for $\theta$, then, we can search for $k$. 

\begin{itemize}
    \item Step 1: Determine if the wound is healed. We realize that for small $k$ (e.g. $k = 0.1$), we can try all the possibilities of $\theta$. As mentioned in the previous section, $\theta$ determines whether the wound is healed. Thus, with all possibilities of $\theta$ in range $\left(0.0, 1.0\right)$ combining with $k = 0.1$, we can find the best combination that yields the lowest loss function. From this, we can determine  the $\theta$ and mark the wound as healed or unhealed as a whole. 
    \item Step 2: We now know if the wound is healed or unhealed as a whole. We can now start searching for optimal $\theta$ and $k$. If the wound is healed, for fixed $k$, we start at $\theta = 0.1$ and increase $\theta$ until the loss function is minimized. If the wound is unhealed as a whole, for fixed $k$, we start at $\theta = 0.9$ and decrease $\theta$ to the optimal value. We repeat this process for low values of $k$ until we obtain the optimal value of $\theta$. From experience, we see that the optimal $\theta$ converges rapidly to a fixed number $\overline{\theta}$.
    \item Step 3: Fixing $\theta$ to be $\overline{\theta}$, we search the optimal $k$ in the specified range.
\end{itemize}

The algorithm described above is designed to be computationally efficient. It utilizes a variant of gradient search to avoid having to evaluate the loss function through all values of $(\theta, k)$.


\chapter{Accuracy of the wound healing model}
\label{ch:results}
\section{Deep Learning Model Accuracy}
We now examine the accuracy among three deep learning models --- U-Net, Res-Unet and Mask-RCNN. We divide our dataset into training, validation and testing sets where our training set accounts for $70\%$ of the dataset. First, we consider U-Net and Residual U-Net models because of their applications in the segmentation field. 
We compare the accuracy and the loss between U-Net and Residual U-Net (see \cite{RFB15a} and \cite{zhang2018road}). Although performing well on the training set, the Residual U-Net model performs poorly on the test set for this segmentation problem. Furthermore, we can easily see that the Residual U-Net model is over-fitting. The reason is the limitations of the labeled data that we have. Meanwhile, the Residual U-Net is more complex than the regular U-Net model when it includes the U-Net model and the residual neural network. 
\begin{figure}[!htbp]
    \centering
    \includegraphics[scale = 0.3]{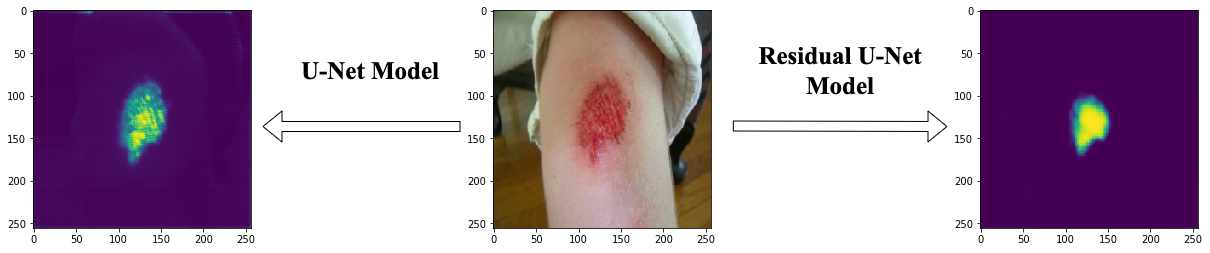}
    \caption{Comparing between U-Net Model and Residual U-Net Model on a picture from Valley Occupational Medical Center.}
    \label{fig:u-net-compare}
\end{figure}
\begin{figure}[!htbp]
\begin{minipage}{.5\textwidth}
  \centering
  \includegraphics[scale=0.4]{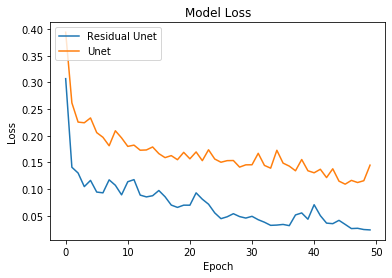}
  \label{fig:test1}
\end{minipage}%
\begin{minipage}{.5\textwidth}
  \centering
  \includegraphics[scale=0.4]{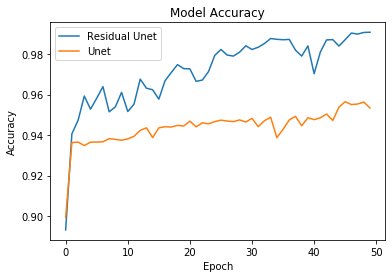}
  \label{fig:test2}
\end{minipage}
\caption{The loss and accuracy of U-Net and Residual U-Net models on the training set.}
\label{fig:test5}
\end{figure}

\FloatBarrier
Moreover, when adapting the Mask-RCNN from detectron2, we achieve a significant improvement in accuracy compared to U-Net Model and Residual U-Net Model on the same test set. Thus, we adopt Mask-RCNN as the baseline model for our full process.
    \begin{figure}
    \centering
\includegraphics[width=0.6\textwidth]{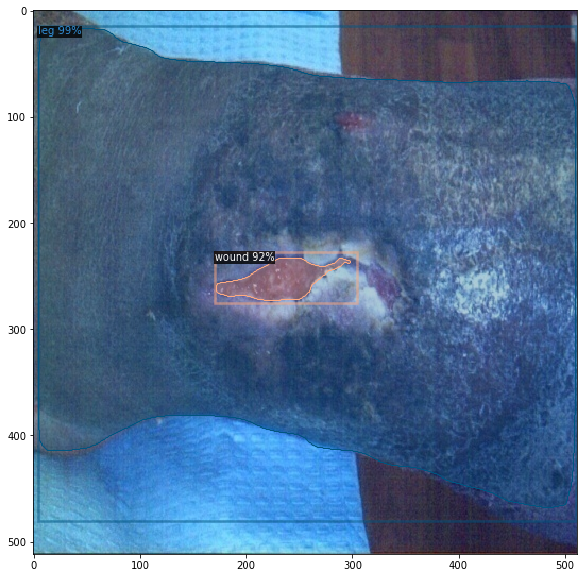}
    \caption{The Mask R-CNN Result Example}
    \label{fig:my_label}
    \end{figure}
    \newpage
\section{Wound Parameter Model Accuracy}
\begin{figure}[!htbp]
    \centering
    \includegraphics[scale=0.8]{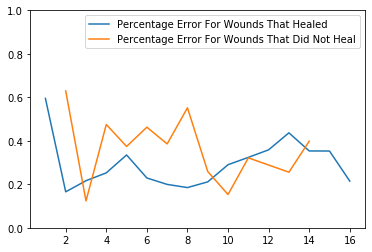}
    \caption{The Average Percentage of Error by Week}
    \label{fig:result}
\end{figure}

By working on the parameter searching method in Section \ref{parametersearching} with the loss function, we divide the dataset which contains 23 patients' data into 2 categories - healing and non-healing sets. By considering each patient separately, we then average the percentage error and represent them by weeks. 
\begin{equation*}
    \text{Percentage  Error} = \frac{\text{predicted wound area} - \text{actual wound area}}{\text{initial wound area}},
\end{equation*} We achieve the results as shown in Figure \ref{fig:result}. The result verifies the accuracy of the models. Because there are many factors that affect the wound healing process for individual patients, we believe that these error percentages are low and show that our method provides a promising model for wound healing.
\FloatBarrier


\chapter{Concluding Remarks}
\label{ch:conclusion}
The whole process of wound healing modelling using the modified-ADI and Deep Learning method provides a new way to model wound behavior. 
We proved the stability of the Euler Method and ADI Method for approximation problems and we found that \textbf{the ADI method} is more efficient than the Euler Method in the stability. Thus, we have adapted it in the whole modelling process.
Moreover, by analyzing different Deep Learning models for segmentation, we determined that Mask-RCNN is a very effective model for this part of the process.
In the future, if more comprehensive wound data is available, we could potentially refine the model to have better accuracy.

Another future variant of this model could potentially identify the phases of wound healing as described in \cite{sherratt1990models}. This opens the door to further investigation.


\begin{appendices}

\addtocontents{toc}{\protect\renewcommand{\protect\cftchappresnum}{\appendixname\space}}
\addtocontents{toc}{\protect\renewcommand{\protect\cftchapnumwidth}{6em}}


\chapter{Stability Analysis of Alternating Direction Implicit}
\label{app:app-A}
The stability of \textbf{ADI Method} can be analyzed by von Neumann method. The two dimensional form of the discrete Fourier Series is: 

\begin{eqnarray}
u^n_{a,b} = \sum^{I-1}_{p=0}\sum^{J-1}_{q=0} C^n_{p,q}e^{2\pi i(pa/I+qb/J)}
\end{eqnarray}

Then substituting this function into \eqref{adi3}, letting $e^{2\pi i(pa/I + qb/J)} = w_{a,b}$ and $w_{a-1,b} = w_{a,b}e^{-2\pi i/I}$, similar for $w_{a+1,b}, w_{a,b-1},w_{a,b+1}$, we have 
\begin{eqnarray}
\sum^{I-1}_{p=0}\sum^{J-1}_{q=0} C^{n+\frac{1}{2}}_{p,q}w_{a,b} &=& \sum^{I-1}_{p=0}\sum^{J-1}_{q=0} C^{n}_{p,q}w_{a,b} + r_x(\sum^{I-1}_{p=0}\sum^{J-1}_{q=0} C^{n+\frac{1}{2}}_{p,q}w_{a-1,b} \nonumber \\
&-& 2\sum^{I-1}_{p=0}\sum^{J-1}_{q=0} C^{n+\frac{1}{2}}_{p,q}w_{a,b} + \sum^{I-1}_{p=0}\sum^{J-1}_{q=0} C^{n+\frac{1}{2}}_{p,q}w_{a+1,b}) \nonumber \\
&+& r_y(\sum^{I-1}_{p=0}\sum^{J-1}_{q=0} C^{n}_{p,q}w_{a,b-1} - 2\sum^{I-1}_{p=0}\sum^{J-1}_{q=0} C^{n}_{p,q}w_{a,b} + \sum^{I-1}_{p=0}\sum^{J-1}_{q=0} C^{n}_{p,q}w_{a,b+1}) \nonumber\\
\sum^{I-1}_{p=0}\sum^{J-1}_{q=0} C^{n+\frac{1}{2}}_{p,q}w_{a,b} &=& \sum^{I-1}_{p=0}\sum^{J-1}_{q=0} C^{n}_{p,q}w_{a,b} + r_x(\sum^{I-1}_{p=0}\sum^{J-1}_{q=0} C^{n+\frac{1}{2}}_{p,q}w_{a,b}e^{-2\pi i /I} \nonumber \\
&-& 2\sum^{I-1}_{p=0}\sum^{J-1}_{q=0} C^{n+\frac{1}{2}}_{p,q}w_{a,b} + \sum^{I-1}_{p=0}\sum^{J-1}_{q=0} C^{n+\frac{1}{2}}_{p,q}w_{a,b}e^{2\pi i /I}) \nonumber \\
&+& r_y(\sum^{I-1}_{p=0}\sum^{J-1}_{q=0} C^{n}_{p,q}w_{a,b}e^{-2\pi i /J} - 2\sum^{I-1}_{p=0}\sum^{J-1}_{q=0} C^{n}_{p,q}w_{a,b} + \sum^{I-1}_{p=0}\sum^{J-1}_{q=0} C^{n}_{p,q}w_{a,b}e^{2\pi i /I}) \nonumber\\
\end{eqnarray}{}
Consider 

\begin{eqnarray}
C_{p,q}^{n+\frac{1}{2}} = C^n_{p,q} + r_x(C_{p,q}^{n+\frac{1}{2}}(e^{-2\pi i/I} - 2 + e^{2\pi i/I})) +
                          r_y(C_{p,q}^{n}(e^{-2\pi i/J} - 2 + e^{2\pi i/J}))\nonumber\\ 
\Longleftrightarrow{} C^{n+\frac{1}{2}}_{p,q}(1-r_x(e^{-2\pi i/I} - 2 + e^{2\pi i/I})) = C^{n}(1 + r_y(e^{-2\pi i/J} - 2 + e^{2\pi i/J}))\nonumber\\
\Longleftrightarrow{}
C^{n+\frac{1}{2}}_{p,q} = C^{n}_{p,q}\frac{1 + r_y(e^{-2\pi i/J} - 2 + e^{2\pi i/J})}{1-r_x(e^{-2\pi i/I} - 2 + e^{2\pi i/I})}.
\end{eqnarray}{}
Substitute $C_{p,q}^{n+\frac{1}{2}}$ into \eqref{adi2} and using a similar process, we have
\begin{eqnarray}
C^{n+1}_{p,q} = C^{n+\frac{1}{2}}_{p,q}\frac{1+r_x(e^{-2\pi i/I} - 2 + e^{2\pi i/I})}{1 - r_y(e^{-2\pi i/J} - 2 + e^{2\pi i/J})}
\nonumber\\
\Longleftrightarrow{}
C^{n+1}_{p,q} =C^{n}_{p,q}(\frac{1 + r_y(e^{-2\pi i/J} - 2 + e^{2\pi i/J})}{1-r_x(e^{-2\pi i/I} - 2 + e^{2\pi i/I})})(\frac{1+r_x(e^{-2\pi i/I} - 2 + e^{2\pi i/I})}{1 - r_y(e^{-2\pi i/J} - 2 + e^{2\pi i/J})})
\label{A.4}\end{eqnarray}{}
with $M_{p,q} =\left(\frac{1 + r_y(e^{-2\pi i/J} - 2 + e^{2\pi i/J})}{1-r_x(e^{-2\pi i/I} - 2 + e^{2\pi i/I})}\right)\left(\frac{1+r_x(e^{-2\pi i/I} - 2 + e^{2\pi i/I})}{1 - r_y(e^{-2\pi i/J} - 2 + e^{2\pi i/J})}\right)$.
Consider $r_x(e^{-2\pi i/I} - 2 + e^{2\pi i/I})$
we have $e^{i\theta} = \cos\theta + i\sin\theta$, hence \begin{eqnarray*}
r_x(e^{-2\pi i/I} - 2 + e^{2\pi i/I}) &=& r_x(\cos(-2\pi /I)  + i\sin(-2\pi /I) - 2 + \cos(2\pi /I) + i\sin(2\pi /I)) \\ &=& r_x(2\cos(2\pi /I) - 2) 
\end{eqnarray*}
which is always negative or 0 since $\cos(2\pi / I) \leq 1$.
Similarly 
$r_y(e^{-2\pi i/J} - 2 + e^{2\pi i/J}) \leq 0$ as well
Let $A = r_x(e^{-2\pi i/I} - 2 + e^{2\pi i/I})$ and $B = r_y(e^{-2\pi i/J} - 2 + e^{2\pi i/J}) \leq 1$, hence we have \eqref{A.4} will become:
\begin{eqnarray*}
C_{p,q}^{n+1} &=& C^n_{p,q}\left(\frac{1+B}{1-A}\right)\left(\frac{1+A}{1-B}\right) \\
\Longleftrightarrow 
C_{p,q}^{n+1} &=& C^n_{p,q}\frac{1+A+B+AB}{(1-A)(1-B)} \\&=& C^n_{p,q}\left(1 + \frac{2A+2B}{(1-A)(1-B)}\right)
\\&=&C^n_{p,q}\left(1 + \frac{2(A+B)}{(1-A)(1-B)}\right) 
\end{eqnarray*}{}
As above, we see $A,B \leq 0$ and thus $\frac{2(A+B)}{(1-A)(1-B)} \leq 0$. Then $M_{p,q} = \left(1 + \frac{2(A+B)}{(1-A)(1-B)}\right) \leq 1$.
\begin{eqnarray}
C^n_{p,q} = (M_{p,q})^nC^0_{p,q}
\end{eqnarray}
where $C^0_{p,q}$ is a Fourier component of the initial data and $M_{p,q}$ is the amplification factor. Also, due to the 1-D analysis, we have that $|M_{p,q}| \leq 1$ for all $\Delta t, \Delta x, \Delta y$. Then the \textbf{ADI method} is unconditionally stable.

\end{appendices}


\begin{singlespace}  
	\setlength\bibitemsep{\baselineskip}  
	\printbibliography[title={References}]

\end{singlespace}

\addcontentsline{toc}{chapter}{References}  

\end{document}//